\documentclass[reprint,amsmath,amssymb,aps,prb, superscriptaddress, showkeys]{revtex4-2}

\usepackage{graphicx}
\usepackage{float}
\usepackage{dcolumn}
\usepackage{xcolor}
\usepackage{soul}
\usepackage[version=4]{mhchem}
\usepackage{layouts}
\usepackage{orcidlink}

\begin{document}

\title{High-Throughput Screening of Phononic Bandgap\\ in Experimentally Realized Materials}

\author{Mohammad Bagheri, \orcidlink{0000-0002-3501-2779}}
\email{mohammad.m.bagheri@jyu.fi}
\affiliation{Nanoscience Center, Department of Physics, University of Jyv\"askyl\"a, 40014 Jyv\"askyl\"a, Finland}

\author{Kameyab Raza Abidi, \orcidlink{0000-0001-5215-8465}}
\affiliation{Nanoscience Center, Department of Physics, University of Jyv\"askyl\"a, 40014 Jyv\"askyl\"a, Finland}

\author{Pekka Koskinen, \orcidlink{0000-0001-7711-3562}}
\email{pekka.j.koskinen@jyu.fi}
\affiliation{Nanoscience Center, Department of Physics, University of Jyv\"askyl\"a, 40014 Jyv\"askyl\"a, Finland}
\date{\today}

\begin{abstract}
Materials with a phonon band gap offer a unique mechanism to restrict phonon scattering, making them promising candidates for developing functional materials for targeted thermal applications. Unlocking this mechanism across broad material classes requires systematic exploration, but first-principles methods and experiments are too expensive.
Here, we used the universal machine-learning interatomic potential to quantify phonons and thermodynamic properties. We validated the accuracy of machine-learning-calculated results by comparing them with density-functional theory.
We screened $154{,}879$ materials and performed phonon calculations for $42{,}347$ thermodynamically stable and experimentally observed compounds. We then performed phonon band gap analysis for dynamically stable materials and identified $11{,}176$ materials with a phononic gap, including $10{,}960$ materials with optical-optical gaps and $270$ with acoustic-optical gaps.
We also analyzed the lattice thermal conductivity of materials with phonon band gaps to identify candidates with acoustic-optical gaps and low thermal conductivity. 
Finally, by solving the linearized phonon Boltzmann equation from first principles, we demonstrate low room-temperature lattice thermal conductivities in three Li-metal binaries.
\end{abstract}

\keywords{phonon band gaps, thermal properties, machine-learning interatomic potentials, density-functional theory}

\maketitle

\section{Introduction}
Lattice vibrations (phonons) in crystals govern the thermal and mechanical behavior of materials.
When vibrational propagation is forbidden over a specific frequency range, a phonon band gap (PBG) emerges, altering energy transport \cite{Argaman_2019, Arya_2025}.
In particular, an acoustic-optical (AO) PBG suppresses scattering channels between acoustic and optical modes, providing an effective mechanism to tune lattice thermal conductivity \cite{Lindsay_2013, Bin_2025}. As a result, materials with an AO PBG have wide applications in thermoelectric materials, sensors, and radio frequency filters \cite{Qian2021, Sledzinska_2020, Guo_2018, Florez2022, Bolin_2015}. 

Extensive research has focused on PBG engineering to tune vibrational, mechanical, and thermal properties via external strain, defect engineering, or chemical doping \cite{Sgouros_2012, Yang_2014, Jiang_2014, Yu_2021, Hong_2019}. However, systematically exploring these configurations experimentally remains challenging and costly.

Computational screening is a natural alternative, yet calculating phonons from first-principles calculations for thousands of materials is computationally demanding. To overcome this bottleneck, universal machine-learning interatomic potentials (UMLIPs) have emerged as powerful tools that significantly accelerate simulations while maintaining density-functional theory (DFT)-level accuracy \cite{Deringer19_AM, Behler2021_review, Jinnouchi2025_review, benedini2025, Kristian2025, sandwich}.
UMLIPs can reproduce accurate energies, forces, and stresses of bulk materials. They have also been shown to reliably predict derivative properties, such as formation energies \cite{Riebesell2024matbench}, defect formation energies \cite{Berger2025, interface}, phonon dispersion curves \cite{alexandria}, and force constants \cite{MLFCDimen}. By accurately capturing phonons, UMLIPs should enable PBG identification across materials.

Therefore, in this article, we screened $154{,}879$ materials from the Materials Project (MP) database \cite{mp} and used UMLIP to perform phonon calculations for experimentally observed compounds. First, we benchmarked ML-calculated PBGs and thermodynamic properties, including Helmholtz free energy, entropy, and constant-volume heat capacity (C$_V$), against DFT using an existing phonon database \cite{Togo2015, alexandria}. Among $42{,}347$ calculated compounds, we identified $11{,}176$ materials with PBG larger than $1$~THz ($\approx 4.14$~meV), including $10{,}960$ of optical-optical (OO) and $270$ of AO. 
We also collected lattice thermal conductivity data for materials with PBGs from the AFLOW materials database \cite{AFLOW} to identify candidates with AO gaps and low thermal conductivity. Finally, we evaluate the lattice thermal conductivity of three Li-metal binaries, LiPt, LiTl, and LiBi, as representative low AO gap materials, by solving the linearized phonon Boltzmann equation using third-order anharmonic force constants, revealing how their structural and vibrational properties produce low thermal conductivity at room temperature.

\section{Methods}

\begin{figure*}
\begin{center}
  \includegraphics[width=\textwidth]{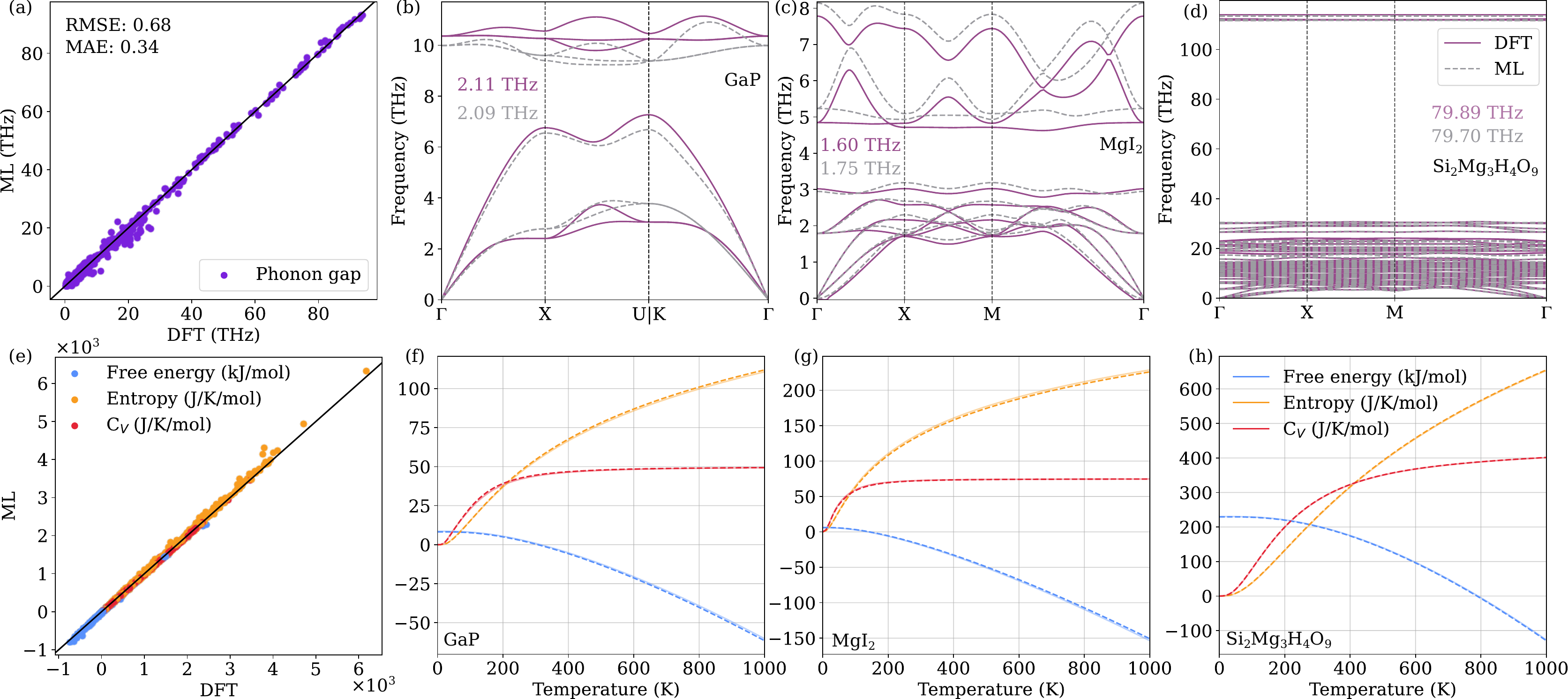}
\end{center}
\caption{\label{fig:benchmark}
\textbf{Comparison of machine-learning (ML) and density-functional theory (DFT) for materials in the Phonon database \cite{Togo2015, alexandria}}. (a) Comparing calculated phonon band gaps using ML (MatterSim) and DFT (PBE). The root-mean-square error (RMSE) and mean absolute error (MAE) are indicated. (b)--(d) Comparing phonon dispersion curves for \ce{GaP} (mp-2490), \ce{MgI2} (mp-23205), and \ce{Si2Mg3H4O9} (mp-23764), respectively. Solid lines are DFT, and dashed lines are ML results.  (e)  Comparison of calculated thermal properties, including Helmholtz free energy (kJ/mol, blue), entropy (J/K/mol, yellow), and constant-volume heat capacity (C$_V$) at $300$~K (J/K/mol, red), using ML and DFT. (f)--(h) Comparing calculated thermal properties for \ce{GaP}, \ce{MgI2}, and \ce{Si2Mg3H4O9}, respectively.}
\end{figure*}

\subsection{Computational methods}
All density-functional theory calculations were performed using VASP (Vienna Ab initio Simulation Package) software \cite{Kresse1996, PhysRevB.54.11169}, together with the projector-augmented wave method \cite{PhysRevB.50.17953}. The plane wave cutoff was set to $520$ eV. The Brillouin zone of the unit cell is sampled by a $\Gamma$-centered k-point mesh whose density is defined by $R_k=20$, which in VASP determines the subdivisions $N_1$, $N_2$, and $N_3$ along the reciprocal lattice vectors $b_1$, $b_2$, and $b_3$, respectively, via $N_i = {\rm max}(1, R_k |b_i| + 0.5)$ and rounded to an integer. The energy convergence criterion between consequential steps is set to $10^{-8}$~eV for the self-consistent loop and $10^{-6}$~eV for ionic relaxation. Structures were considered to be relaxed when no force exceeded $10^{-4}$~eV/\AA.

The second-order and third-order force constants (FCs) are calculated by the finite displacement method, using a displacement of $0.01$~\AA.
The second-order FCs in the harmonic approximation, thermodynamic properties including Helmholtz free energy, entropy, and constant-volume heat capacity (C$_v$), and the phonon calculations were done using the Phonopy package \cite{phonopy1, phonopy2}. 
The third-order (anharmonic) FCs and the lattice thermal conductivity ($\kappa$) are calculated using the single-mode relaxation time approximation (RTA) and the linearized phonon Boltzmann equation by the Phono3py package \cite{phono3py, phonopy-phono3py-JPCM}. The lattice thermal conductivity is reported as the directional average of the diagonal tensor components ($\kappa_{xx}, \kappa_{yy}$, and $\kappa_{zz}$).

The supercell sizes used to determine the force constants were chosen to keep the number of atoms less than or equal to $150$, while keeping the point groups of the Bravais lattices of the conventional unit cell and the supercell the same. 
A \textbf{q}-mesh of $22 \times 22 \times 22$ is used for the calculation of lattice thermal conductivity of Li-metal binaries.

The machine-learning (ML) calculations were performed using MatterSim with pre-trained model v$1.0.0-5$M \cite{MatterSim}.
ASE \cite{ase-paper} was used, with Python, to construct the calculation workflow. The FIRE \cite{FIRE} algorithm was used to relax materials to forces below $1$~meV/\AA.
The crystal structures and their properties were extracted using MP-API \cite{MP_API} and Python materials genomics (Pymatgen) \cite{pymatgen}.

The lattice thermal conductivity values of materials with a phonon band gap were extracted from AFLOW \cite{AFLOW} using the dedicated RESTful Application Programming Interface (API) \cite{AFLOW2}.

\begin{figure*}
\begin{center}
  \includegraphics[width=\textwidth]{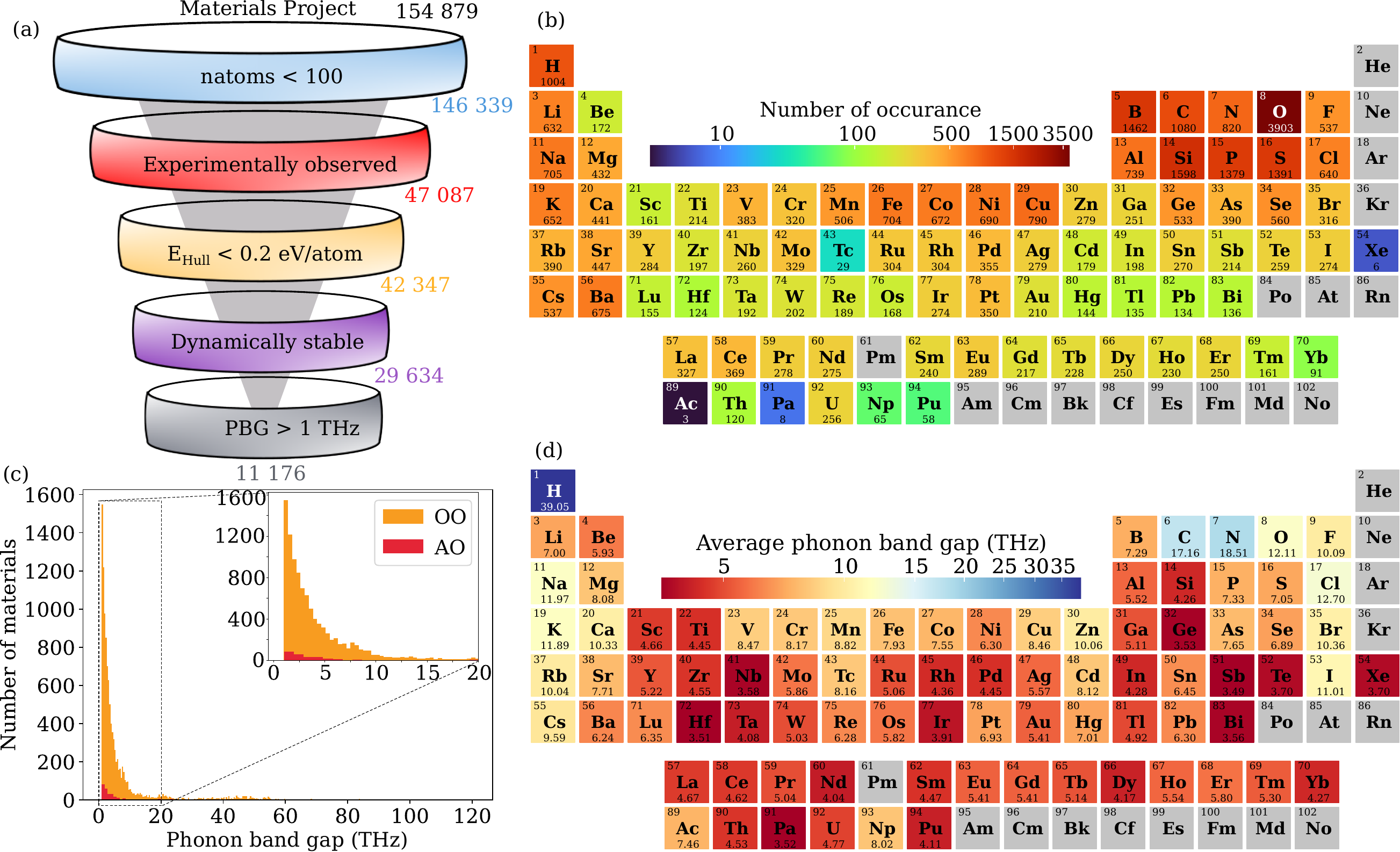}
\end{center}
\caption{\label{fig:dataset}
\textbf{High-throughput calculations}. (a) Screening procedure of the Materials Project (MP) database. The number of materials in each step is indicated. (b) Heatmap of the number of occurrences of periodic table elements in the screened dataset on a logarithmic scale. (c) Histogram of the number of phononic band gap materials in the dataset as a function of optical--optical (OO) and acoustic--optical (AO). (d) Heat map of the average phonon band gap of each element in the screened dataset.}
\end{figure*}

\section{Results and discussion}

\subsection{Benchmark and validation}

We first benchmark the UMLIP against DFT calculations to verify the accuracy of the phonon dispersion and PBGs. Specifically, we employ MatterSim \cite{MatterSim} as our ML calculator, as it has demonstrated reliable predictions of vibrational properties, including force constants and phonons \cite{alexandria, MLFCDimen}.
We used the recalculated version of the Phonon database \cite{Togo2015, alexandria}, including $\approx 10$K materials originally calculated using the PBESol functional, with the PBE functional as the DFT reference for a fair comparison. To ensure structural validity, we restricted our dataset to the $3558$ materials common to both MatterSim and the Phonon database that are dynamically stable (no imaginary frequencies $>0.1$~THz).
We performed phonon calculations and identified $2420$ materials with a PBG, of which $2367$ have an OO PBG and $48$ have an AO PBG. The PBGs from ML are pretty close to the values from DFT, especially for large PBG $>40$~THz, with an overall root mean square error (RMSE) of $0.68$~THz and a mean absolute error (MAE) of $0.34$~THz, which is sufficient for PBG analysis (Fig.~\ref{fig:benchmark}a).

We note that all gap types were similar between ML and DFT, except for five materials (mp-23818, mp-10182, mp-7581, mp-567636, and mp-12558) (see Supplemental Material (SM) Fig.~S1 \cite{SM}).
Differences in phonon gap classification in the above materials arise because full-Brillouin-zone gap detection depends strictly on global band envelopes, where minor ML frequency shifts ($0.2-0.4$~THz) can artificially open, close, or reorder narrow band gaps. Filtering for gaps larger than $1$~THz [which is much larger than our minimum gap threshold ($\delta_{\mathrm{gap}}$)] eliminates false detections from model noise, which should be considered in selection rules.

We selected three materials to demonstrate the phonon dispersion curves comparison: \ce{GaP} (mp-2490) and \ce{MgI2} (mp-23205), which are two well-known AO and OO PBG materials, and \ce{Si2Mg3H4O9} (mp-23764), which is a material with a very large PBG ($>79$~THz). 
The calculated phonon dispersions and PBG for these materials show good agreement with previously reported literature values \cite{Fu_1999, Zhu_2014, Zhang_2024, Togo2015, Bagheri_2023}.
Although there may be some frequency shifts in ML modes, especially in optical bands, the PBGs are similar (Fig.~\ref{fig:benchmark}b--d).

High-fidelity phonon data enable a rigorous assessment of the thermal properties of materials in the benchmark dataset. Therefore, we calculated Helmholtz free energy, entropy, and constant-volume heat capacity (C$_v$) for the benchmark dataset. Comparing the ML data with DFT yielded very close results (Fig.~\ref{fig:benchmark}e); in particular, the ML-predicted and DFT-calculated C$_v$ are almost identical, with an RMSE of $3.66$~J/K/mol and an MAE of $1.59$~J/K/mol. The errors in predicted free energy are larger, with an RMSE of $9.94$~kJ/mol and an MAE of $4.81$~kJ/mol. Similarly, entropy errors show an RMSE of $28.50$~J/K/mol and an MAE of $13.53$~J/K/mol.

\subsection{High-throughput calculations}

We proceed to perform phonon calculations and identify materials with PBG in the MP database. We applied a screening criterion before starting the high-throughput calculations (Fig.~\ref{fig:dataset}a). We limited our study to materials with fewer than $100$ atoms in the conventional unit cell and to experimentally observed materials based on MP entries. We also removed materials with energy above convex hull ($E_{Hull}$) more than $0.2$ eV/atom. Next, we performed phonon calculations for $42{,}347$ materials and found that $29{,}634$ are dynamically stable (no imaginary frequencies $>-0.1$~THz). Finally, we analyzed the phonon band structures and identified materials with PBGs exceeding $1$~THz ($\approx 4.14$~meV). The number of occurrences of each element in $11{,}176$ materials with PBGs shows that oxides are the most common, with an $\approx 35$\% share of materials in our dataset, followed by Si, B, S, P, and H (Fig.~\ref{fig:dataset}b). Notably, alkali and alkali-earth metals comprised a large group in our dataset.

We identified $270$ AO PBG materials, and more than $98$\% of our dataset ($10{,}960$ materials) consists of OO PBG materials (Fig.~\ref{fig:dataset}c). Most PBGs are below $20$~THz, and the maximum PBG is $23.92$~THz for AO and $120.59$~THz for OO. 
As expected, the average PBG across elements in the periodic table reveals a pronounced dependence on atomic mass, where light elements exhibit significantly larger band gaps than heavier elements (Fig.~\ref{fig:dataset}d). Hydrogen dominates the distribution, with a maximum of $39.05$~THz, followed by first-row $p$-block elements such as nitrogen ($18.51$~THz) and carbon ($17.16$~THz), driven by the large frequency separation between lightweight optical vibrations and heavier or acoustic modes. In contrast, the average PBG reduces for lighter alkali and alkaline-earth metals (e.g., Na at $11.97$~THz and K at $11.89$~THz), while the vast majority of $d$-block transition metals and post-transition metals fall into the lower regime below $6$~THz (reaching minima such as Sb at $3.49$~THz, Hf at $3.51$~THz, and Ge at $3.53$~THz).

\begin{figure}
  \centering
  \includegraphics[width=\linewidth]{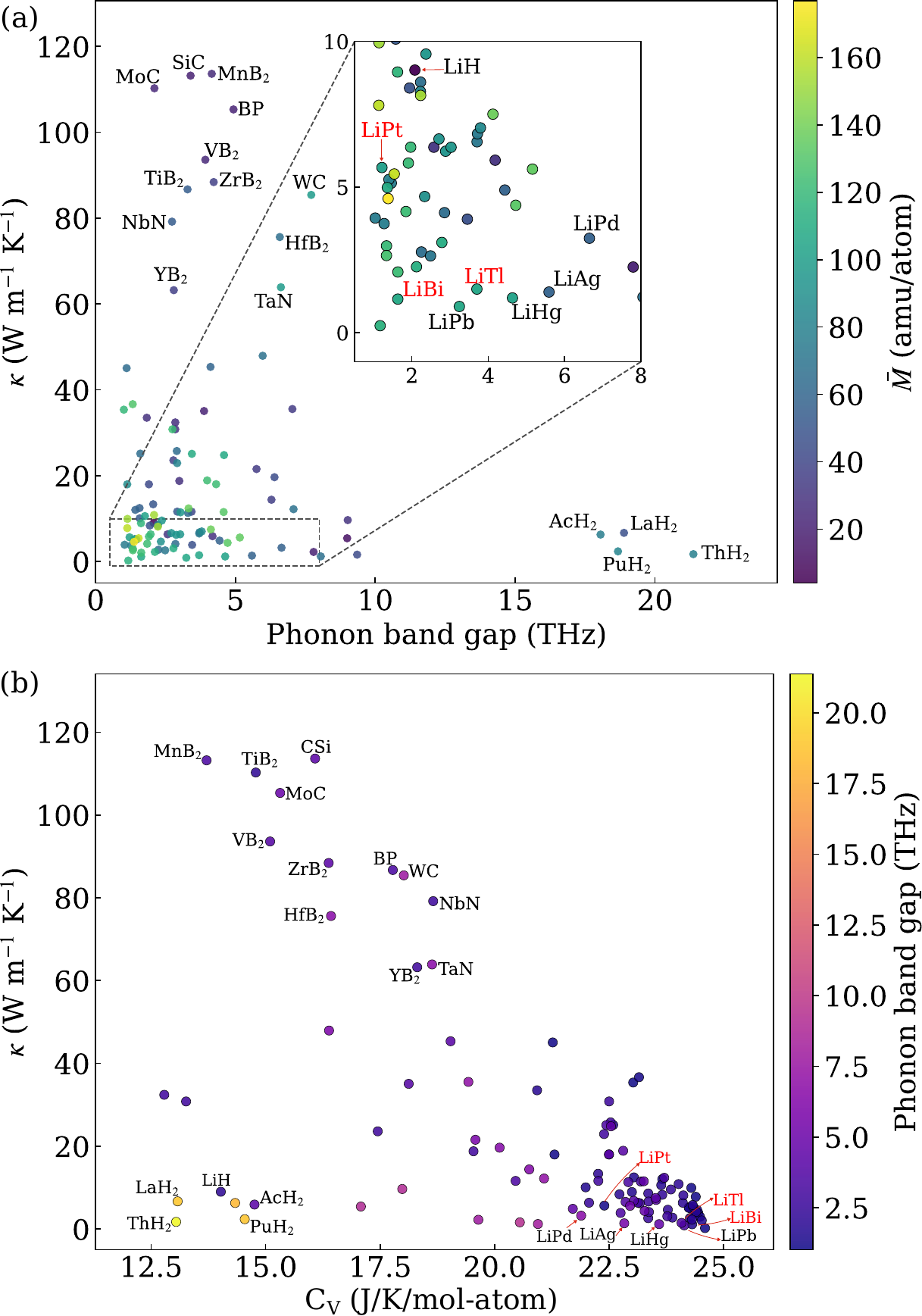}
  \caption{ \textbf{Room-temperature lattice thermal conductivity ($\kappa$) correlations across AO PBG materials.} (a) $\kappa$ (extracted from AFLOW \cite{AFLOW}) as a function of PBG. The values are color-coded based on average mass. Li-metal binary candidates for further investigation are marked in red. (b) $\kappa$ (extracted from AFLOW \cite{AFLOW}) as a function of molar heat capacity ($C_V$) at $300$~K, color-coded by PBG.}
  \label{fig:ao}
\end{figure}

\begin{figure*}
  \centering
  \includegraphics[width=\textwidth]{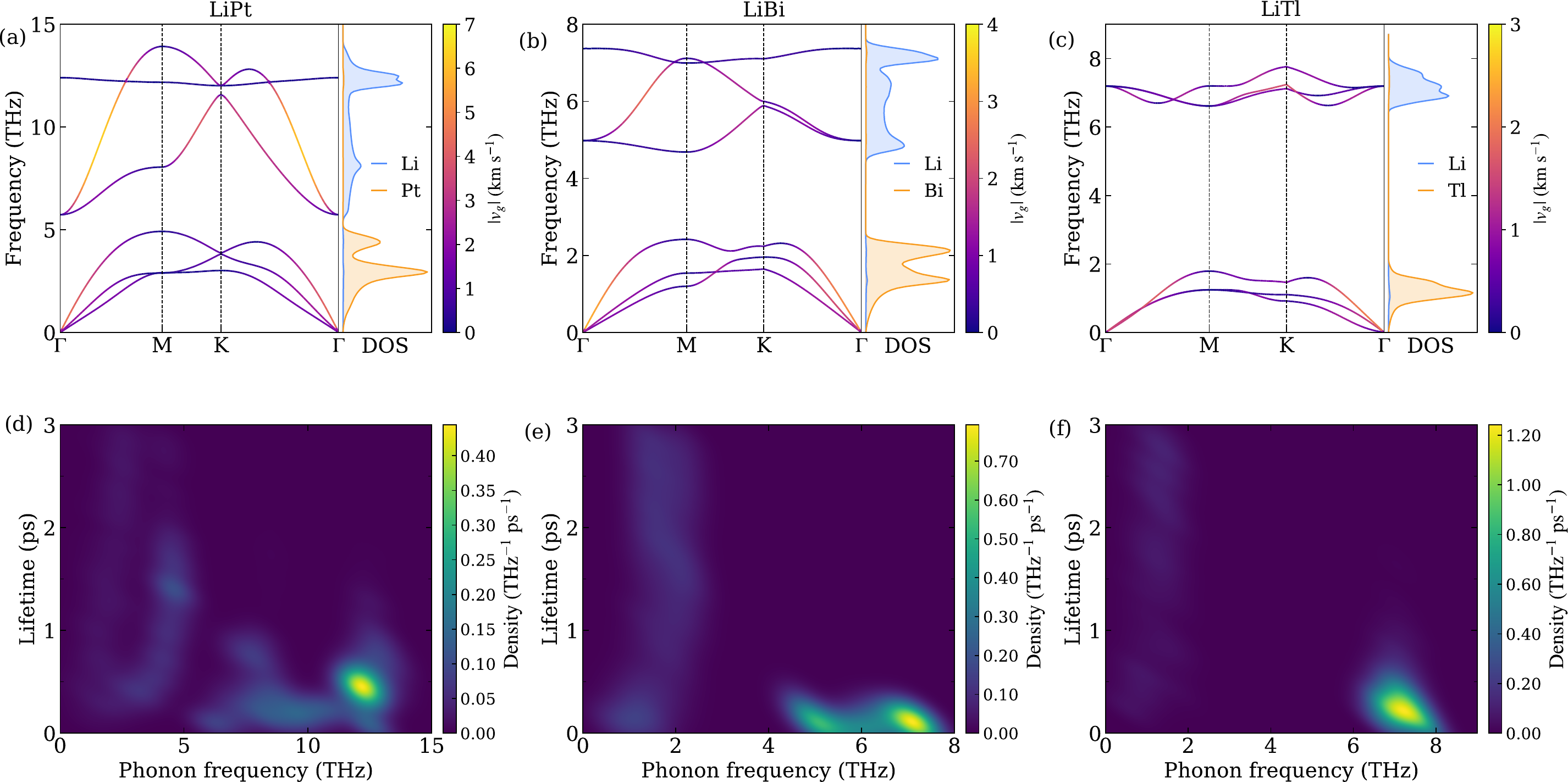}
  \caption{ \textbf{Li--metal binaries}. Phonon dispersion curves and phonon projected density of states of  (a) LiPt (mp-11807), (b) LiBi (mp-22902), and (c) LiTl (mp-934). Bands are colored based on group velocities ($|v_g|$).  Phonon lifetimes of  (d) LiPt, (e) LiBi, and (f) LiTl at $300$~K. The color represents the phonon density. Brighter colors indicate higher phonon density (THz$^{-1}$\ ps$^{-1}$).
  }
  \label{fig:li}
\end{figure*}

\subsection{Phonon band gap and thermal conductivity}
Here, we focus on the thermodynamic and lattice thermal conductivities of materials exhibiting PBGs. By cross-referencing MP and AFLOW entries via unique Inorganic Crystal Structure Database (ICSD) identifiers \cite{icsd}, we retrieved the room-temperature lattice thermal conductivity ($\kappa$) for $993$ compounds across both AO and OO PBG systems (SM Tables~S1 and S2 \cite{SM}). To evaluate vibrational-mode population alongside heat transport, we also evaluated the molar heat capacity at constant volume ($C_V$) at $300$~K.

The room-temperature lattice thermal conductivities were extracted from the AFLOW materials database. In AFLOW, $\kappa$ was computed using the Automatic GIBBS Library (AGL) coupled with the Automatic Elasticity Library (AEL) ~\cite{Cormac_2017}. While this model omits microscopic scattering details and lacks absolute quantitative accuracy, its strong correlation with experimental trends makes it an efficient screening descriptor. Promising candidates can subsequently be modeled using full third-order anharmonic Boltzmann transport equation (BTE) simulations for quantitative accuracy.

The impact of an AO PBG on $\kappa$ depends on which phonon scattering channels are restricted or preserved by the dispersion, so materials exhibiting an AO PBG naturally occupy diverse thermal transport regimes.
In AO PBG materials in our dataset, $\kappa$ spans more than two orders of magnitude across moderate band gaps ($<5$~THz). We identify prominent high-$\kappa$ conductors, including SiC, \ce{MnB2}, \ce{MoC}, and \ce{BP}, which benefit from small average atomic masses ($\bar{M}$), stiff interatomic bonding, and high acoustic group velocities (Fig.~\ref{fig:ao}a). Conversely, systems displaying exceptionally wide gaps ($>15$~THz), such as actinide and lanthanide hydrides (\ce{AcH2}, \ce{LaH2}, \ce{PuH2}, and \ce{ThH2}), exhibit low lattice thermal conductivities ($\kappa <5$). In these hydrides, the substantial gap is driven predominantly by the light hydrogen atom, which shifts optical branches to frequencies well above $20$~THz while leaving the heavy-metal acoustic branches low-lying and densely scattered.
The same trend holds for OO materials, while the low-PBG materials show a smaller $\kappa$ than the AO materials (SM Fig.~S2 \cite{SM}). This makes AO PBG more interesting for further studies. 

Comparing $\kappa$ against $C_V$ (Fig.~\ref{fig:ao}b) highlights a clear decoupling between equilibrium thermodynamic heat capacity and non-equilibrium transport. High-gap hydrides and light-element ceramics display substantially reduced heat capacities ($C_V \sim 12.5$--15~J/K/mol-atom), reflecting the quantum freeze-out of high-frequency optical modes at 300~K. In contrast, most materials cluster near the classical Dulong--Petit limit ($C_V \approx 3R \approx 24.9$~J/K/mol-atom), yet they span a broad conductivity spectrum. A similar distribution appears for OO materials (SM Fig.~S2 \cite{SM}), although their low-PBG members generally yield lower peak $\kappa$ values than their AO counterparts, making AO systems particularly versatile for thermal transport engineering.

We can identify promising low-$\kappa$ candidates by zooming in around $\kappa < 10$~W~m$^{-1}$~K$^{-1}$ (Fig.~\ref{fig:ao}a), where most of the materials are governed by high-mass systems ($\bar{M} > 100$~amu/atom). Among them, we highlight a group of lithium--metal binaries (LiPt, LiPd, LiBi, LiTl, LiPb, LiHg, and LiAg). The extreme mass disparity between light lithium ($M_{\text{Li}} \approx 6.94$~amu) and heavy $p$- and $d$-block partners ($M \sim 106-209$~amu) opens an AO gap while heavily depressing acoustic sound velocities. Consequently, these binary systems consistently show low thermal conductivities ($\kappa < 7$~W~m$^{-1}$~K$^{-1}$, with several dipping below $2$~W~m$^{-1}$~K$^{-1}$), establishing them as ideal target systems for computationally demanding higher-order phonon scattering and detailed lattice thermal conductivity investigations.

\subsection{Origins of low thermal conductivity}

To uncover the mechanism behind the low thermal conductivity identified in the previous section, we performed lattice thermal conductivity calculations using DFT for three representative binary systems: LiPt (mp-11807), LiBi (mp-22902), and LiTl (mp-934). 
The calculated phonon dispersion curves and atom-projected density of states (DOS) show clear AO gaps across all three materials (Figs.~\ref{fig:li}a--c). Because of the extreme atomic mass ratio, the vibrational spectrum is sharply partitioned: low-frequency acoustic modes below $2-5$~THz stem almost entirely from the heavy metal atoms (Pt, Bi, Tl), whereas high-frequency optical branches are governed by Li atoms. The heavy-metal sublattice compresses the acoustic branches, producing very low sound velocities; in particular, acoustic group velocities in LiBi and LiTl remain well below 2~km~s$^{-1}$ (see SM Fig.~S3 \cite{SM} for more detail).
Strong anharmonic scattering is evident across all branches, as demonstrated by the evaluated phonon lifetimes at $300$~K (Figs.~\ref{fig:li}d--f). Mode lifetimes are strongly suppressed, with most falling below $1$~ps and almost none exceeding $2$~ps. This pronounced phase-space scattering, combined with flattened dispersions, keeps the phonon mean free paths short.

\begin{figure}
  \centering
  \includegraphics[width=0.85\linewidth]{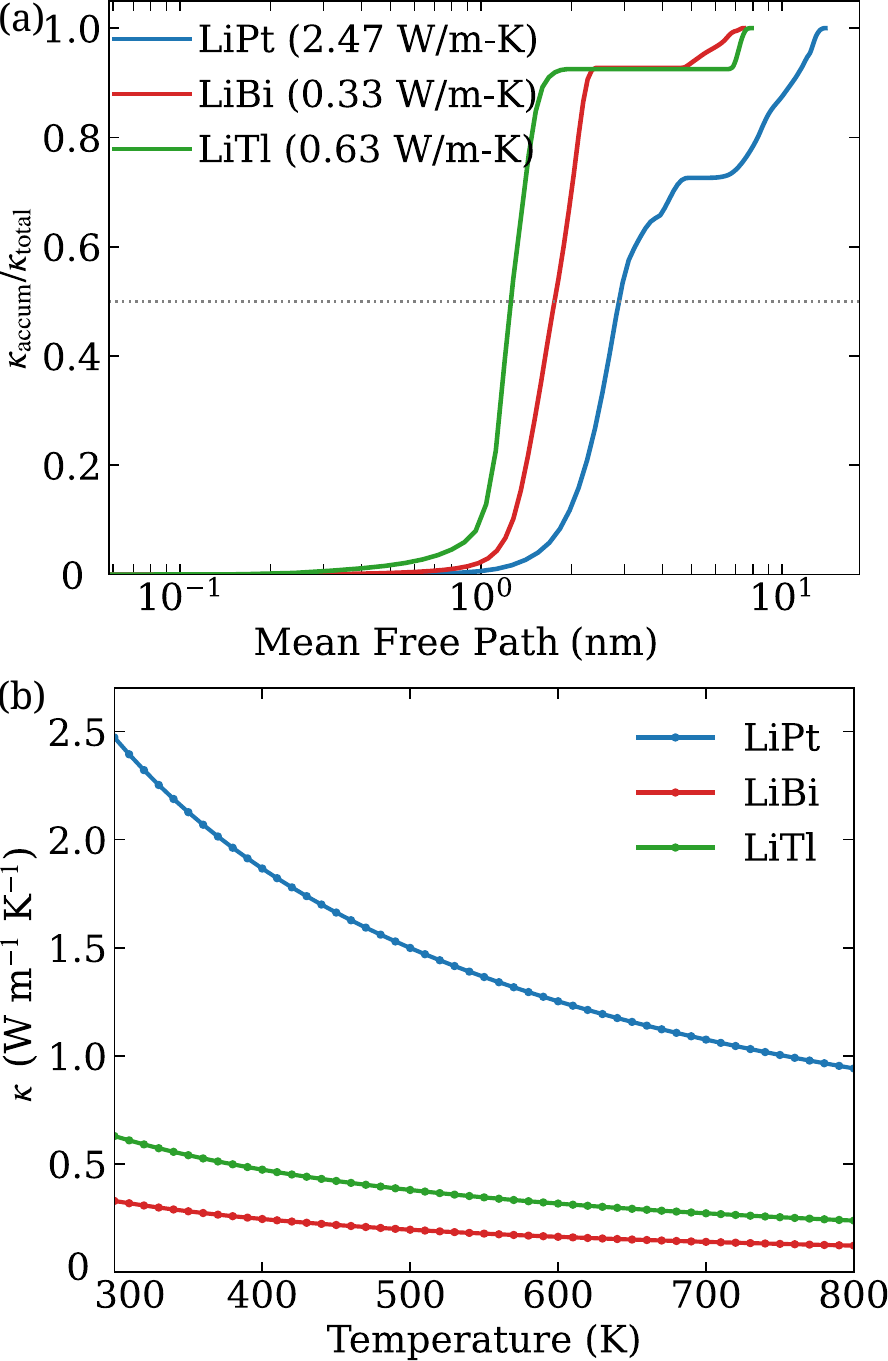}
  \caption{\textbf{Lattice thermal conductivity and mean free path accumulation of selected Li-metal binaries.}  (a) Normalized cumulative lattice thermal conductivity ($\kappa_{\text{accum}}/\kappa_{\text{total}}$) at $300$~K plotted against phonon mean free path. The horizontal dotted line marks the $50$\% accumulation threshold. (b) Calculated lattice thermal conductivity ($\kappa$) as a function of temperature from $300$ to $800$~K.
  }
  \label{fig:kappa}
\end{figure}

Spectral decomposition of the cumulative thermal conductivity ($\kappa_{\text{accum}}/\kappa_{\text{total}}$) reveals how heat is partitioned across the vibrational spectrum (SM Fig.~S4 \cite{SM}). Across the AO gap, the curve flattens because no vibrational states are available to conduct heat. While heavy-metal acoustic phonons carry $\approx 60$\% of the heat, the optical Li branches contribute a considerable $\approx 35-40$\% of total $\kappa$ across all three systems. Furthermore, cumulative tracking against mean free path demonstrates that $50$\% of the heat is carried by phonons with mean free paths below $2.8$~nm in LiPt and below $1.8$~nm in LiBi and LiTl, with over $90$\% accumulated below $8$~nm (Fig.~\ref{fig:kappa}a).

Consequently, the temperature-dependent lattice thermal conductivity ($\kappa$) confirms intrinsic low values across the entire temperature range (Fig.~\ref{fig:kappa}b). At $300$~K, $\kappa$ reaches $2.47$~W~m$^{-1}$~K$^{-1}$ for LiPt, $0.63$~W~m$^{-1}$~K$^{-1}$ for LiTl, and drops to an exceptionally low $0.33$~W~m$^{-1}$~K$^{-1}$ for LiBi. With increasing temperature up to $800$~K, $\kappa$ follows typical $1/T$ Umklapp decay \cite{Broido_2007, Lindsay_2008}, declining to glass-like limits \cite{Cahill_1992, Simoncelli2019} of $0.95$~W~m$^{-1}$~K$^{-1}$ in LiPt and below $0.25$~W~m$^{-1}$~K$^{-1}$ in both LiTl and LiBi. These nanoscopic mean free paths and glass-like thermal conductivities demonstrate that combining large mass-mismatch AO gaps with heavy-element acoustic suppression provides a robust mechanism for designing crystalline materials with low thermal conductivities for thermoelectric and thermal-barrier applications.

\section{Conclusions}
In conclusion, we demonstrated that MatterSim predicts phonon band gaps (PBGs) and thermodynamic properties with first-principles-level accuracy.
We performed phonon calculations for $42{,}347$ thermodynamically stable and experimentally observed materials. 
We then analyzed PBGs for dynamically stable materials and identified $11{,}176$ materials with a gap larger than $1$~THz, including $10{,}960$ with optical-optical (OO) gaps and $270$ with acoustic-optical (AO) gaps.
We also studied the lattice thermal conductivity of materials with PBGs and found candidates with AO and OO gaps with low thermal conductivity. 
Finally, by solving the linearized phonon Boltzmann equation from first principles, we demonstrate low room-temperature lattice thermal conductivities in three Li-metal binaries: LiPt, LiTl, and LiBi.
Overall, our systematic computational approach maps an extensive library of PBG materials with tailored thermal conductivities, broadening their potential across functional applications.

\section*{Acknowledgments}
We acknowledge the Jane and Aatos Erkko Foundation for funding (project EcoMet). We also thank the Finnish Grid and Cloud Infrastructure (FGCI) and CSC—IT Center for Science for computational resources. We acknowledge the EuroHPC Joint Undertaking for awarding this project access to the EuroHPC supercomputer LUMI, hosted by CSC (Finland) and the LUMI consortium through a EuroHPC Regular Access call.

\section*{Data availability}
All data supporting the findings have been deposited in a JSON database at \href{https://doi.org/10.5281/zenodo.22868227}{\url{https://doi.org/10.5281/zenodo.22868227}}.

\section*{Appendices}
\section*{Appendix A: Phonon band gap detection}
Phonon frequencies $\omega(\mathbf{q}\nu)$ were obtained by diagonalizing the dynamical matrix at each wave vector $\mathbf{q}$, sampled on a $\Gamma$-centered uniform mesh of $N_1 \times N_2 \times N_3$ points in the Brillouin zone.
Here $\nu = 1, \ldots, N_{\mathrm{b}}$ indices the branches in ascending frequency order, $N_{\mathrm{b}} = 3s$ for a computational cell containing $s$ atoms, and $\mathbf{q}$ runs over all mesh points. For each branch $\nu$, the sampled frequency envelope is a closed interval as
\begin{align}  
    \nonumber I_\nu \;=\; \bigl[\,\omega_\nu^{\min},\;\omega_\nu^{\max}\bigr],\\
    \omega_\nu^{\min/\max} \;\equiv\; \min/\!\max_{\mathbf{q}}\;\omega(\mathbf{q}\nu).
    \label{eq:branch_interval}
\end{align}

To construct the covered spectral regions, the interval lower bounds are set to zero ($\omega_\nu^{\min} \rightarrow \max(0, \omega_\nu^{\min})$), and any branch whose maximum frequency is $\leq 0$ is discarded.
The set of valid intervals $\{I_\nu\}$ is sorted by their lower bounds, and any two consecutive intervals whose separation is smaller than a merging tolerance $\delta_{\mathrm{m}}$ are fused into a single interval. This prevents numerical near-degeneracies at avoided crossings from producing spurious gaps.
A phonon gap is then defined as any frequency interval not covered by the merged intervals, provided its width exceeds a minimum threshold $\delta_{\mathrm{gap}}$.
We apply the gap criterion directly to the branch envelopes on the reciprocal $\mathbf{q}$-point mesh.

Gaps are classified by the branches they separate. The $n_{\mathrm{ac}}~(=3)$ lowest branches are designated acoustic, and the remaining branches are optical. This assignment is validated at the $\Gamma$-point by requiring $|\omega(\Gamma\nu)| \leq \delta_\Gamma$ for all $\nu \leq n_{\mathrm{ac}}$ and $|\omega(\Gamma, n_{\mathrm{ac}}+1)| > \delta_\Gamma$; structures failing this check are excluded from rigorous gap classification. The AO gap width is then
\begin{equation}
    \Delta_{\mathrm{ao}}
    \;=\;
    \omega_{\mathrm{opt}}^{\min} - \omega_{\mathrm{ac}}^{\max},
    \label{eq:ao_gap}
\end{equation}
where $\omega_{\mathrm{ac}}^{\max} = \max_{\mathbf{q},\,\nu \leq n_{\mathrm{ac}}}\omega(\mathbf{q}\nu)$ is the upper edge of the acoustic frequency range and $\omega_{\mathrm{opt}}^{\min} = \max\bigl(0, \min_{\mathbf{q},\,\nu > n_{\mathrm{ac}}}\omega(\mathbf{q}\nu)\bigr)$ is the lower edge of the optical frequency range; an AO gap is reported if $\Delta_{\mathrm{ao}} \ge \delta_{\mathrm{gap}}$.
OO gaps are located analogously from the merged envelopes of branches $\nu > n_{\mathrm{ac}}$.
The numerical parameters used throughout are $n_{\mathrm{ac}} = 3$ (fixed by translational invariance), the numerical tolerances $\delta_{\Gamma} = \delta_{\mathrm{imag}} = 0.10$~THz (set safely above typical acoustic sum rule residuals), the merging tolerance $\delta_{\mathrm{m}} = 0.02$~THz (accounting for finite-mesh artifacts), the minimum gap threshold $\delta_{\mathrm{gap}} = 0.30$~THz (ensuring robustness against mesh refinement and anharmonic broadening), and uniform mesh dimensions were set to $N_i = 81$ for structures with fewer than $40$ atoms, and $N_i = 41$ otherwise.\\

\section*{Appendix B: Phonon thermodynamic properties}

Within the harmonic approximation, the vibrational contributions to the Helmholtz free energy, entropy, and heat capacity were computed from the phonon frequencies $\omega(\mathbf{q}\nu)$, obtained by diagonalizing the dynamical matrix at each wave vector $\mathbf{q}$ \cite{phonopy1, phonopy2, phonopy_thermal}. The Helmholtz free energy per unit cell is given by
\begin{equation}
\begin{split}
    F(T) = \varphi &+ \frac{1}{2}\sum_{\mathbf{q}\nu} \hbar\omega(\mathbf{q}\nu) \\
    &+ k_{\mathrm{B}}T \sum_{\mathbf{q}\nu} \ln\Bigl[1 - \exp\Bigl(-\frac{\hbar\omega(\mathbf{q}\nu)}{k_{\mathrm{B}}T}\Bigr)\Bigr],
\end{split}
\label{eq:free_energy}
\end{equation}
where $\varphi$ is the static-lattice electronic energy per unit cell, $\omega(\mathbf{q}\nu)$ is the angular frequency of the phonon mode at wave vector $\mathbf{q}$ and branch index $\nu$, $\hbar$ is the reduced Planck constant, $k_{\mathrm{B}}$ is the Boltzmann constant, and $T$ is the temperature. The summations run over all phonon modes sampled across the reciprocal $\mathbf{q}$-point mesh in the first Brillouin zone. In Eq.~\eqref{eq:free_energy}, the second term represents the temperature-independent zero-point energy, while the third term accounts for finite-temperature thermal excitations.

The vibrational entropy, derived via $S = -(\partial F/\partial T)_V$, is expressed as
\begin{equation}
\begin{split}
    S(T) = &\frac{1}{2T}\sum_{\mathbf{q}\nu} \hbar\omega(\mathbf{q}\nu) \coth\Bigl(\frac{\hbar\omega(\mathbf{q}\nu)}{2k_{\mathrm{B}}T}\Bigr) \\
    &- k_{\mathrm{B}}\sum_{\mathbf{q}\nu} \ln\Bigl[2\sinh\Bigl(\frac{\hbar\omega(\mathbf{q}\nu)}{2k_{\mathrm{B}}T}\Bigr)\Bigr].
\end{split}
\label{eq:entropy}
\end{equation}

The heat capacity is calculated as
\begin{equation}
    C_V(T) = k_{\mathrm{B}} \sum_{\mathbf{q}\nu} 
    \frac{\left(\frac{\hbar\omega(\mathbf{q}\nu)}{k_{\mathrm{B}}T}\right)^2 \exp\left[\frac{\hbar\omega(\mathbf{q}\nu)}{k_{\mathrm{B}}T}\right]}
    {\left\{\exp\left[\frac{\hbar\omega(\mathbf{q}\nu)}{k_{\mathrm{B}}T}\right] - 1\right\}^2}.
\label{eq:heat_capacity}
\end{equation}

\bibliography{refs}

@article{Argaman_2019,
  title = {Phonon band gaps in the IV-VI monochalcogenides},
  author = {Argaman, Uri and Abutbul, Ran E. and Golan, Yuval and Makov, Guy},
  journal = {Phys. Rev. B},
  volume = {100},
  issue = {5},
  pages = {054104},
  numpages = {12},
  year = {2019},
  publisher = {American Physical Society},
  doi = {10.1103/PhysRevB.100.054104}
}

@article{Arya_2025,
    author = {Arya, Chaitanya and Trautvetter, Johannes and Sojo-Gordillo, Jose M. and Kaur, Yashpreet and Zannier, Valentina and Nigro, Arianna and Beltram, Fabio and Albrigi, Tommaso and Ruiz-Caridad, Alicia and Sorba, Lucia and Rurali, Riccardo and Zardo, Ilaria},
    title = {Phonon Interference Effects in {G}a{A}s-{G}a{P} Superlattice Nanowires},
    journal = {ACS Nano},
    volume = {19},
    number = {50},
    pages = {42013-42021},
    year = {2025},
    issn = {1936-0851},
    doi = {10.1021/acsnano.5c10312}
}

@article{Yu_2021,
    author = {Yu, Xiaoxia and Hong, Jiawang},
    title = {Absence of phonon gap driven ultralow lattice thermal conductivity in half-Heusler {LuNiBi}},
    journal = {	J. Mater. Chem. C},
    volume = {9},
    number = {36},
    pages = {12420-12425},
    year = {2021},
    issn = {2050-7526},
    doi = {10.1039/d1tc02819g}
}

@article{Hong_2019,
    author = {Hong, Min and Wang, Yuan and Feng, Tianli and Sun, Qiang and Xu, Shengduo and Matsumura, Syo and Pantelides, Sokrates T. and Zou, Jin and Chen, Zhi-Gang},
    title = {Strong Phonon–Phonon Interactions Securing Extraordinary Thermoelectric {Ge1–xSbxTe} with {Zn}-Alloying-Induced Band Alignment},
    journal = {J. Am. Chem. Soc.},
    volume = {141},
    number = {4},
    pages = {1742-1748},
    year = {2019},
    issn = {0002-7863},
    doi = {10.1021/jacs.8b12624}
}

@article{Yang_2014,
    author = {Yang, Lina and Yang, Nuo and Li, Baowen},
    title = {Extreme Low Thermal Conductivity in Nanoscale 3{D} {Si} Phononic Crystal with Spherical Pores},
    journal = {Nano Lett.},
    volume = {14},
    number = {4},
    pages = {1734-1738},
    year = {2014},
    issn = {1530-6984},
    doi = {10.1021/nl403750s}
}

@article{MLFCDimen,
      title={Massive Discovery of Low-Dimensional Materials from Universal Computational Strategy}, 
      author={Mohammad Bagheri and Ethan Berger and Hannu-Pekka Komsa and Pekka Koskinen},
    journal = {Chem. Mater.},
     volume = {38},
     number = {5},
     pages = {2395–-2402},
     year = {2026},
     doi = {https://doi.org/10.1021/acs.chemmater.5c03151}
}

@article{interface,
author ={Bagheri, Mohammad and Koskinen, Pekka},
title  ={Lateral graphene-metallene interfaces at the nanoscale},
journal  ={Nanoscale},
year  ={2026},
volume  ={18},
issue  ={1},
pages  ={188-196},
publisher  ={The Royal Society of Chemistry},
doi  ={10.1039/D5NR02770E}
}

@article{Qian2021,
  title = {Phonon-engineered extreme thermal conductivity materials},
  volume = {20},
  ISSN = {1476-4660},
  DOI = {10.1038/s41563-021-00918-3},
  number = {9},
  journal = {Nat. Mater.},
  publisher = {Springer Science and Business Media LLC},
  author = {Qian,  Xin and Zhou,  Jiawei and Chen,  Gang},
  year = {2021},
  pages = {1188–1202}
}

@article{Deringer19_AM,
author = {Deringer, Volker L. and Caro, Miguel A. and Cs\'{a}nyi, G\'{a }bor},
title = {Machine Learning Interatomic Potentials as Emerging Tools for Materials Science},
journal = {Adv. Mater.},
volume = {31},
number = {46},
pages = {1902765},
doi = {https://doi.org/10.1002/adma.201902765},
year = {2019}
}

@article{Behler2021_review,
author = {Behler, J{\"o}rg},
title = {Four Generations of High-Dimensional Neural Network Potentials},
journal = {Chem. Rev.},
volume = {121},
number = {16},
pages = {10037-10072},
year = {2021},
doi = {10.1021/acs.chemrev.0c00868}
}

@article{Jinnouchi2025_review,
author = {Jinnouchi, Ryosuke and Minami, Saori},
title = {Machine Learning Force Fields in Electrochemistry: From Fundamentals to Applications},
journal = {ACS Nano},
volume = {19},
number = {25},
pages = {22600-22644},
year = {2025},
doi = {10.1021/acsnano.5c05553}
}

@article{Riebesell2024matbench,
  title = {A Framework to Evaluate Machine Learning Crystal Stability Predictions},
  volume = {7},
  ISSN = {2522-5839},
  DOI = {10.1038/s42256-025-01055-1},
  number = {6},
  journal = {Nat. Mach. Intell.},
  publisher = {Springer Science and Business Media LLC},
  author = {Riebesell,  Janosh and Goodall,  Rhys E. A. and Benner,  Philipp and Chiang,  Yuan and Deng,  Bowen and Ceder,  Gerbrand and Asta,  Mark and Lee,  Alpha A. and Jain,  Anubhav and Persson,  Kristin A.},
  year = {2025},
  pages = {836–847}
}

@article{Berger2025,
author = {Berger, Ethan and Bagheri, Mohammad and Komsa, Hannu-Pekka},
title = {Screening of Material Defects using Universal Machine-Learning Interatomic Potentials},
journal = {Small},
volume = {21},
number = {37},
pages = {e03956},
year = {2025},
doi = {https://doi.org/10.1002/smll.202503956}
}

@article{alexandria,
  title = {Universal Machine Learning Interatomic Potentials Are Ready for Phonons},
  volume = {11},
  issn = {2057-3960},
  doi = {10.1038/s41524-025-01650-1},
  pages = {178},
  journal = {npj Comput. Mater.},
  publisher = {Springer Science and Business Media LLC},
  author = {Loew,  Antoine and Sun,  Dewen and Wang,  Hai-Chen and Botti,  Silvana and Marques,  Miguel A. L.},
  year = {2025}
}

@article{benedini2025,
      title={Universal Machine Learning Potential for Systems with Reduced Dimensionality}, 
      author={Giulio Benedini and Antoine Loew and Matti Hellstrom and Silvana Botti and Miguel A. L. Marques},
    doi = {10.1088/3050-287X/ae1208},
    year = {2025},
    publisher = {IOP Publishing},
    volume = {1},
    number = {2},
    pages = {025005},
    journal = {AI Sci.}
}

@article{Kristian2025,
  title = {Dispersion-Corrected Machine Learning Potentials for 2{D} Van der Waals Materials},
  author = {Sauer, Mikkel Ohm and Lyngby, Peder Meisner and Thygesen, Kristian Sommer},
  journal = {Phys. Rev. Mater.},
  volume = {9},
  issue = {7},
  pages = {074007},
  numpages = {11},
  year = {2025},
  publisher = {American Physical Society},
  doi = {10.1103/cl8c-8f1f}
}

@article{Kresse1996,
author = {Kresse, G. and Furthm{\"{u}}ller, J.},
doi = {https://doi.org/10.1016/0927-0256(96)00008-0},
journal = {Comput. Mater. Sci.},
number = {1},
pages = {15--50},
title = {Efficiency of Ab-Initio Total Energy Calculations for Metals and Semiconductors using a Plane-Wave Basis Set},
volume = {6},
year = {1996}
}

@article{PhysRevB.54.11169,
  title = {Efficient Iterative Schemes for Ab Initio Total-Energy Calculations using a Plane-Wave Basis Set},
  author = {Kresse, G. and Furthm\"uller, J.},
  journal = {Phys. Rev. B},
  volume = {54},
  issue = {16},
  pages = {11169--11186},
  numpages = {0},
  year = {1996},
  publisher = {American Physical Society},
  doi = {10.1103/PhysRevB.54.11169}
}

@article{PhysRevB.50.17953,
  title = {Projector Augmented-Wave Method},
  author = {Bl\"ochl, P. E.},
  journal = {Phys. Rev. B},
  volume = {50},
  issue = {24},
  pages = {17953--17979},
  numpages = {0},
  year = {1994},
  doi = {10.1103/PhysRevB.50.17953},
  publisher = {American Physical Society}
}

@article{pymatgen,
title = {Python Materials Genomics (pymatgen): A Robust, Open-Source Python Library for Materials Analysis},
journal = {Comput. Mater. Sci.},
volume = {68},
pages = {314-319},
year = {2013},
doi = {https://doi.org/10.1016/j.commatsci.2012.10.028},
author = {Shyue Ping Ong and William Davidson Richards and Anubhav Jain and Geoffroy Hautier and Michael Kocher and Shreyas Cholia and Dan Gunter and Vincent L. Chevrier and Kristin A. Persson and Gerbrand Ceder},
}

@article{MP_API,
	doi = {https://doi.org/10.1016/j.commatsci.2014.10.037},
	year = 2015,
	publisher = {Elsevier {BV}},
	volume = {97},
	pages = {209-215},
	author = {Shyue Ping Ong and Shreyas Cholia and Anubhav Jain and Miriam Brafman and Dan Gunter and Gerbrand Ceder and Kristin A. Persson},
	title = {The Materials Application Programming Interface ({API}): A Simple, Flexible and Efficient {API} for Materials Data Based on Representational State Transfer ({REST}) Principles},
	journal = {Comput. Mater. Sci.}
}

@article{phonopy1,
  author  = {Togo, Atsushi and Chaput, Laurent and Tadano, Terumasa and Tanaka, Isao},
  title   = {Implementation Strategies in Phonopy and Phono3py},
  journal = {J. Phys. Condens. Matter},
  volume  = {35},
  number  = {35},
  pages   = {353001},
  year    = {2023},
  doi     = {10.1088/1361-648X/acd831}
}

@article{phonopy2,
  author  = {Togo, Atsushi},
  title   = {First-Principles Phonon Calculations with Phonopy and Phono3py},
  journal = {J. Phys. Soc. Jpn.},
  volume  = {92},
  number  = {1},
  pages   = {012001},
  year    = {2023},
  doi     = {10.7566/JPSJ.92.012001}
}

@article{phonopy_thermal,
  title = {First-principles phonon calculations of thermal expansion in ${\text{Ti}}_{3}{\text{SiC}}_{2}$, ${\text{Ti}}_{3}{\text{AlC}}_{2}$, and ${\text{Ti}}_{3}{\text{GeC}}_{2}$},
  author = {Togo, Atsushi and Chaput, Laurent and Tanaka, Isao and Hug, Gilles},
  journal = {Phys. Rev. B},
  volume = {81},
  issue = {17},
  pages = {174301},
  numpages = {6},
  year = {2010},
  publisher = {American Physical Society},
  doi = {10.1103/PhysRevB.81.174301}
}

@misc{MatterSim,
      title={MatterSim: A Deep Learning Atomistic Model Across Elements, Temperatures and Pressures}, 
      author={Han Yang and Chenxi Hu and Yichi Zhou and Xixian Liu and Yu Shi and Jielan Li and Guanzhi Li and Zekun Chen and Shuizhou Chen and Claudio Zeni and Matthew Horton and Robert Pinsler and Andrew Fowler and Daniel Zügner and Tian Xie and Jake Smith and Lixin Sun and Qian Wang and Lingyu Kong and Chang Liu and Hongxia Hao and Ziheng Lu},
      year={2024},
      eprint={2405.04967},
      archiveprefix={arXiv},
      url={https://arxiv.org/abs/2405.04967}
}

@article{ase-paper,
  author={Ask Hjorth Larsen and Jens Jørgen Mortensen and Jakob Blomqvist and Ivano E Castelli and Rune Christensen and Marcin
Dułak and Jesper Friis and Michael N Groves and Bjørk Hammer and Cory Hargus and Eric D Hermes and Paul C Jennings and Peter
Bjerre Jensen and James Kermode and John R Kitchin and Esben Leonhard Kolsbjerg and Joseph Kubal and Kristen
Kaasbjerg and Steen Lysgaard and Jón Bergmann Maronsson and Tristan Maxson and Thomas Olsen and Lars Pastewka and Andrew
Peterson and Carsten Rostgaard and Jakob Schiøtz and Ole Schütt and Mikkel Strange and Kristian S Thygesen and Tejs
Vegge and Lasse Vilhelmsen and Michael Walter and Zhenhua Zeng and Karsten W Jacobsen},
  title={The Atomic Simulation Environment—A Python Library for Working with Atoms},
  journal={J. Phys.:Condens. Matter.},
  volume={29},
  number={27},
  pages={273002},
  year={2017}
}

@article{FIRE,
  title = {Structural Relaxation Made Simple},
  author = {Bitzek, Erik and Koskinen, Pekka and G\"ahler, Franz and Moseler, Michael and Gumbsch, Peter},
  journal = {Phys. Rev. Lett.},
  volume = {97},
  issue = {17},
  pages = {170201},
  numpages = {4},
  year = {2006},
  publisher = {American Physical Society},
  doi = {10.1103/PhysRevLett.97.170201}
}

@article{Lindsay_2013,
  title = {First-Principles Determination of Ultrahigh Thermal Conductivity of Boron Arsenide: A Competitor for Diamond?},
  author = {Lindsay, L. and Broido, D. A. and Reinecke, T. L.},
  journal = {Phys. Rev. Lett.},
  volume = {111},
  issue = {2},
  pages = {025901},
  numpages = {5},
  year = {2013},
  publisher = {American Physical Society},
  doi = {10.1103/PhysRevLett.111.025901}
}

@article{Sgouros_2012,
    author = {Sgouros, A. and Sigalas, M. M. and Kalosakas, G. and Papagelis, K. and Papanicolaou, N. I.},
    title = {Phononic band gap engineering in graphene},
    journal = {J. Appl. Phys.},
    volume = {112},
    number = {9},
    pages = {094307},
    year = {2012},
    issn = {0021-8979},
    doi = {10.1063/1.4763479}
}

@article{Jiang_2014,
author ={Jiang, Jin-Wu},
title  ={Phonon bandgap engineering of strained monolayer {MoS2}},
journal  ={Nanoscale},
year  ={2014},
volume  ={6},
issue  ={14},
pages  ={8326-8333},
publisher  ={The Royal Society of Chemistry},
doi  ={10.1039/C4NR00279B}
}

@article{mp,
author = {Jain, Anubhav and Ong, Shyue Ping and Hautier, Geoffroy and Chen, Wei and Richards, William Davidson and Dacek, Stephen and Cholia, Shreyas and Gunter, Dan and Skinner, David and Ceder, Gerbrand and Persson, Kristin a.},
doi = {https://doi.org/10.1063/1.4812323},
journal = {APL Mater.},
number = {1},
pages = {011002},
title = {The Materials Project: A Materials Genome Approach to Accelerating Materials Innovation},
volume = {1},
year = {2013}
}

@article{Togo2015,
title = {First Principles Phonon Calculations in Materials Science},
journal = {Scr. Mater.},
volume = {108},
pages = {1-5},
year = {2015},
doi = {https://doi.org/10.1016/j.scriptamat.2015.07.021},
author = {Atsushi Togo and Isao Tanaka}
}

@article{Sledzinska_2020,
author = {Sledzinska, Marianna and Graczykowski, Bartlomiej and Maire, Jeremie and Chavez-Angel, Emigdio and Sotomayor-Torres, Clivia M. and Alzina, Francesc},
title = {2{D} Phononic Crystals: Progress and Prospects in Hypersound and Thermal Transport Engineering},
journal = {Adv. Funct. Mater.},
volume = {30},
number = {8},
pages = {1904434},
doi = {https://doi.org/10.1002/adfm.201904434},
year = {2020}
}

@article{Guo_2018,
doi = {10.1088/1361-648X/aaab32},
year = {2018},
publisher = {IOP Publishing},
volume = {30},
number = {10},
pages = {105701},
author = {Guo, San-Dong and Liu, Bang-Gui},
title = {Ultrahigh lattice thermal conductivity in topological semimetal {TaN} caused by a large acoustic-optical gap},
journal = {J. Phys.:Condens. Matter.},
}

@article{Florez2022,
  title = {Engineering nanoscale hypersonic phonon transport},
  volume = {17},
  ISSN = {1748-3395},
  DOI = {10.1038/s41565-022-01178-1},
  number = {9},
  journal = {Nat. Nanotechnol.},
  publisher = {Springer Science and Business Media LLC},
  author = {Florez,  O. and Arregui,  G. and Albrechtsen,  M. and Ng,  R. C. and Gomis-Bresco,  J. and Stobbe,  S. and Sotomayor-Torres,  C. M. and García,  P. D.},
  year = {2022},
  pages = {947–951}
}

@article{AFLOW,
title = {aflow.org: A web ecosystem of databases, software and tools},
journal = {Comput. Mater. Sci.},
volume = {216},
pages = {111808},
year = {2023},
issn = {0927-0256},
doi = {https://doi.org/10.1016/j.commatsci.2022.111808},
author = {Marco Esters and Corey Oses and Simon Divilov and Hagen Eckert and Rico Friedrich and David Hicks and Michael J. Mehl and Frisco Rose and Andriy Smolyanyuk and Arrigo Calzolari and Xiomara Campilongo and Cormac Toher and Stefano Curtarolo}
}

@article{AFLOW2,
title = {A {RESTful} {API} for exchanging materials data in the {AFLOWLIB.org} consortium},
journal = {Comput. Mater. Sci.},
volume = {93},
pages = {178-192},
year = {2014},
issn = {0927-0256},
doi = {https://doi.org/10.1016/j.commatsci.2014.05.014},
author = {Richard H. Taylor and Frisco Rose and Cormac Toher and Ohad Levy and Kesong Yang and Marco {Buongiorno Nardelli} and Stefano Curtarolo}
}

@article{phono3py,
  title = {Distributions of phonon lifetimes in Brillouin zones},
  author = {Togo, Atsushi and Chaput, Laurent and Tanaka, Isao},
  journal = {Phys. Rev. B},
  volume = {91},
  issue = {9},
  pages = {094306},
  numpages = {31},
  year = {2015},
  month = {Mar},
  publisher = {American Physical Society},
  doi = {10.1103/PhysRevB.91.094306},
}

@article{phonopy-phono3py-JPCM,
  author  = {Togo, Atsushi and Chaput, Laurent and Tadano, Terumasa and Tanaka, Isao},
  title   = {Implementation strategies in phonopy and phono3py},
  journal = {J. Phys. Condens. Matter.},
  volume  = {35},
  number  = {35},
  pages   = {353001},
  year    = {2023},
  doi     = {10.1088/1361-648X/acd831}
}

@article{Cahill_1992,
  title = {Lower limit to the thermal conductivity of disordered crystals},
  author = {Cahill, David G. and Watson, S. K. and Pohl, R. O.},
  journal = {Phys. Rev. B},
  volume = {46},
  issue = {10},
  pages = {6131--6140},
  numpages = {0},
  year = {1992},
  month = {Sep},
  publisher = {American Physical Society},
  doi = {10.1103/PhysRevB.46.6131},
}

@article{Broido_2007,
    author = {Broido, D. A. and Malorny, M. and Birner, G. and Mingo, Natalio and Stewart, D. A.},
    title = {Intrinsic lattice thermal conductivity of semiconductors from first principles},
    journal = {Appl. Phys. Lett.},
    volume = {91},
    number = {23},
    pages = {231922},
    year = {2007},
    issn = {0003-6951},
    doi = {10.1063/1.2822891}
}

@article{Simoncelli2019,
  title = {Unified theory of thermal transport in crystals and glasses},
  volume = {15},
  ISSN = {1745-2481},
  DOI = {10.1038/s41567-019-0520-x},
  number = {8},
  journal = {Nat. Phys.},
  publisher = {Springer Science and Business Media LLC},
  author = {Simoncelli,  Michele and Marzari,  Nicola and Mauri,  Francesco},
  year = {2019},
  pages = {809–813}
}

@article{Bolin_2015,
  title = {Significant Reduction of Lattice Thermal Conductivity by the Electron-Phonon Interaction in Silicon with High Carrier Concentrations: A First-Principles Study},
  author = {Liao, Bolin and Qiu, Bo and Zhou, Jiawei and Huberman, Samuel and Esfarjani, Keivan and Chen, Gang},
  journal = {Phys. Rev. Lett.},
  volume = {114},
  issue = {11},
  pages = {115901},
  numpages = {6},
  year = {2015},
  publisher = {American Physical Society},
  doi = {10.1103/PhysRevLett.114.115901},
}

@article{Lindsay_2008,
doi = {10.1088/0953-8984/20/16/165209},
year = {2008},
volume = {20},
number = {16},
pages = {165209},
author = {Lindsay, L and Broido, D A},
title = {Three-phonon phase space and lattice thermal conductivity in semiconductors},
journal = {J. Phys. Condens. Matter.}
}

@article{Cormac_2017,
  title = {Combining the {AFLOW GIBBS} and elastic libraries to efficiently and robustly screen thermomechanical properties of solids},
  author = {Toher, Cormac and Oses, Corey and Plata, Jose J. and Hicks, David and Rose, Frisco and Levy, Ohad and de Jong, Maarten and Asta, Mark and Fornari, Marco and Buongiorno Nardelli, Marco and Curtarolo, Stefano},
  journal = {Phys. Rev. Mater.},
  volume = {1},
  issue = {1},
  pages = {015401},
  numpages = {26},
  year = {2017},
  publisher = {American Physical Society},
  doi = {10.1103/PhysRevMaterials.1.015401}
}

@article{Bin_2025,
    author = {Wei, Bin and Zhang, Xiaotian and Li, Wang and Li, Jiale and Li, Yongheng and Gao, Qilong and Hong, Jiawang and Nan, Ce-Wen and Lin, Yuan-Hua},
    title = {Optical phonons on thermal conduction in advanced materials},
    journal = {Appl. Phys. Rev.},
    volume = {12},
    number = {1},
    pages = {011324},
    year = {2025},
    issn = {1931-9401},
    doi = {10.1063/5.0239197}
}

@article{Bagheri_2023,
  doi = {https://doi.org/10.1038/s41597-023-01988-5},
  year = {2023},
  publisher = {Springer Science and Business Media {LLC}},
  volume = {10},
  pages = {80},
  issue = {1},
  author = {M. Bagheri and H.-P. Komsa},
  title = {High-throughput computation of {R}aman spectra from first principles},
  journal = {Sci. Data}
}

@article{Zhang_2024,
    author = {Zhang, Chunwei and Jiang, Xiaobo and Wang, Xiaodan and Cao, Xingan and Zhou, LinZhen and Xing, Yuheng and Xu, Ning},
    title = {The effect of optical-acoustic phonon coupling on the thermal conductivity of 2{D} {M}g{I}2},
    journal = {Phys. Chem. Chem. Phys.},
    volume = {26},
    number = {34},
    pages = {22509-22517},
    year = {2024},
    issn = {1463-9076},
    doi = {10.1039/d4cp02462a},
}

@article{Fu_1999,
  title = {Phonons in {GaP} quantum dots},
  author = {Fu, Huaxiang and Ozoli\ifmmode \mbox{\c{n}}\else \c{n}\fi{}\ifmmode \check{s}\else \v{s}\fi{}, V. and Zunger, Alex},
  journal = {Phys. Rev. B},
  volume = {59},
  issue = {4},
  pages = {2881--2887},
  numpages = {0},
  year = {1999},
  publisher = {American Physical Society},
  doi = {10.1103/PhysRevB.59.2881}
}

@article{Zhu_2014,
    author = {Zhu, Jiajie and Schwingenschl\"ogl, Udo},
    title = {Structural and Electronic Properties of Silicene on {M}g{X}2 ({X = Cl, Br, and I})},
    journal = {ACS Appl. Mater. Interfaces},
    volume = {6},
    number = {14},
    pages = {11675-11681},
    year = {2014},
    issn = {1944-8244},
    doi = {10.1021/am502469m}
}

@article{icsd,
author = {Zagorac, D. and M{\"{u}}ller, H. and Ruehl, S. and Zagorac, J. and Rehme, S.},
title = {Recent developments in the Inorganic Crystal Structure Database: theoretical crystal structure data and related features},
journal = {J. Appl. Crystallogr.},
year = {2019},
volume = {52},
number = {5},
pages = {918--925},
doi = {10.1107/S160057671900997X}
}

@misc{sandwich,
      title={Machine-Learning-Accelerated Metallene Stabilization from High-Throughput Sandwich Modeling}, 
      author={Mohammad Bagheri and Pekka Koskinen},
      year={2026},
      eprint={2608.01779},
      archivePrefix={arXiv},
      primaryClass={cond-mat.mtrl-sci},
      url={https://arxiv.org/abs/2608.01779}, 
}

@misc{SM,
  note = {See Supplemental Material at [URL here] for additional results on benchmarks; thermal conductivity and molar heat capacity of optical--optical phonon band gap material; list of materials with available thermal conductivity and their metadata; supplementary data for thermal conductivity results of Li-metal binaries.}
}

\end{document}